\documentclass[twoside]{dis09}
\usepackage[latin1]{inputenc}
\usepackage[dvips]{graphicx,epsfig,color}
\usepackage{wrapfig,rotating}
\usepackage{amssymb,amsmath,array}

\begin{document}
\title{Charm fragmentation and excited charm and charm-strange mesons at ZEUS}

\author{Leonid Gladilin\thanks{Participation in DIS\,2009 was supported by the Deutsches Elektronen-Synchrotron DESY}
%
~~(on behalf of the ZEUS collaboration)
%
\vspace{.3cm}\\
%
Moscow State University - Scobeltsyn Institute of Nuclear Physics \\
1(2), Leninskie gory, GSP-1, Moscow 119991 - Russia\\
E-mail: gladilin@sinp.msu.ru
}

\maketitle

\begin{abstract}
The charm fragmentation function has been measured in photoproduction
of jets containing $D^{*\pm}$ mesons.
The measured function has been used to extract free parameters
of different fragmentation models.
Measurements of excited charm, $D_1(2420)^0$ and $D_2^*(2460)^0$,
and charm-strange, $D_{s1}(2536)^\pm$, mesons and a search for the
radially excited charm meson, $D^{*\prime}(2640)^\pm$, were
also performed.
The results are compared with those measured previously
and with theoretical expectations~\cite{url}.
\end{abstract}

\section{Introduction}

Charm hadrons were produced copiously in $ep$ collisions
with a centre-of-mass energy of $318\,$GeV
at HERA providing
a means to study charm hadronisation.
During first phase of the HERA operation (1992-2000),
the ZEUS collaboration accumulated data sample corresponding to
an integrated luminosity of $\sim120\,$pb$^{-1}$.
Measurements of the $D^{*+}$, $D^0$, $D^+$ and $\Lambda_c^+$
production\footnote{Hereafter, charge conjugation is implied.}
with the HERA I data
were used to determine charm fragmentation ratios and fractions of $c$
quarks hadronising as a particular charm hadron, $f(c\rightarrow D,\Lambda_c)$,
in earlier ZEUS studies~\cite{ff_php,ff_dis}.
Recent ZEUS measurements of the charm fragmentation function~\cite{zeus_func}
and the production of excited charm and charm-strange mesons~\cite{zeus_excd}
are summarised in this note.

\section{Measurement of charm fragmentation function}

The measurement of the charm fragmentation function in the transition
from a charm quark to a $D^{*+}$ meson
was performed in photoproduction regime with the virtuality
of the exchanged photon $Q^2<1\,$GeV$^2$ and the photon-proton centre-of-mass
energy in the range $130<W<280\,$GeV.
The $D^{*+}$ mesons were reconstructed from the decay chain
$D^{*+}\rightarrow D^0\pi^{+}\rightarrow (K^-\pi^+)\pi^+$
using the mass difference technique.
The $D^{*+}$ meson was included in the jet-finding procedure and was
thereby uniquely associated with one jet only.
The fragmentation variable, $z$, was defined as
$$z=(E+p_{||})^{D^{*+}}/(E+p_{||})^{\rm jet}
\equiv (E+p_{||})^{D^{*+}}/2\,E^{\rm jet},$$
where $p_{||}$ is the longitudinal momentum of the $D^{*+}$ meson
or of the jet
relative to the axis of the jet of energy $E^{\rm jet}$.
The equivalence of $(E+p_{||})^{\rm jet}$ and $2\,E^{\rm jet}$ arises
because the jets were reconstructed as massless objects.
The measurement
of the normalised differential cross section,
$1/\sigma (d\sigma/d z)$,
was performed in the kinematic range
$p_T(D^{*+}) > 2\,$GeV, $|\eta(D^{*+})| < 1.5$,
$E_T^{\rm jet}>9\,$GeV and $|\eta^{\rm jet}| < 2.4$.
The above requirements on $p_T(D^{*+})$ and $E_T^{\rm jet}$
allowed the fragmentation function measurement in the range
$0.16<z<1$.

\begin{figure}[!ht]
\begin{center}
\centerline{\epsfxsize=2.7in\epsfbox{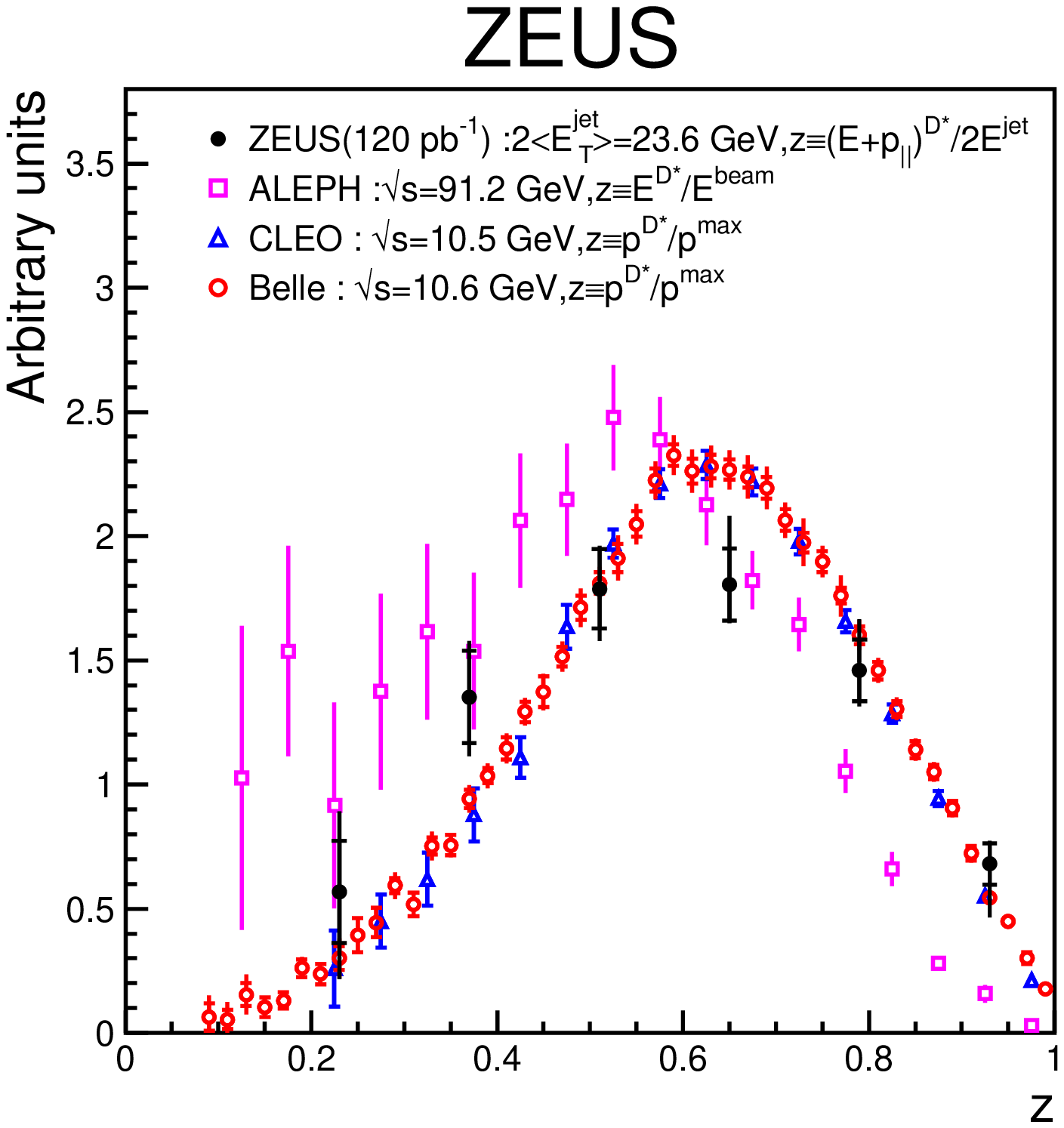}
\epsfxsize=2.7in\epsfbox{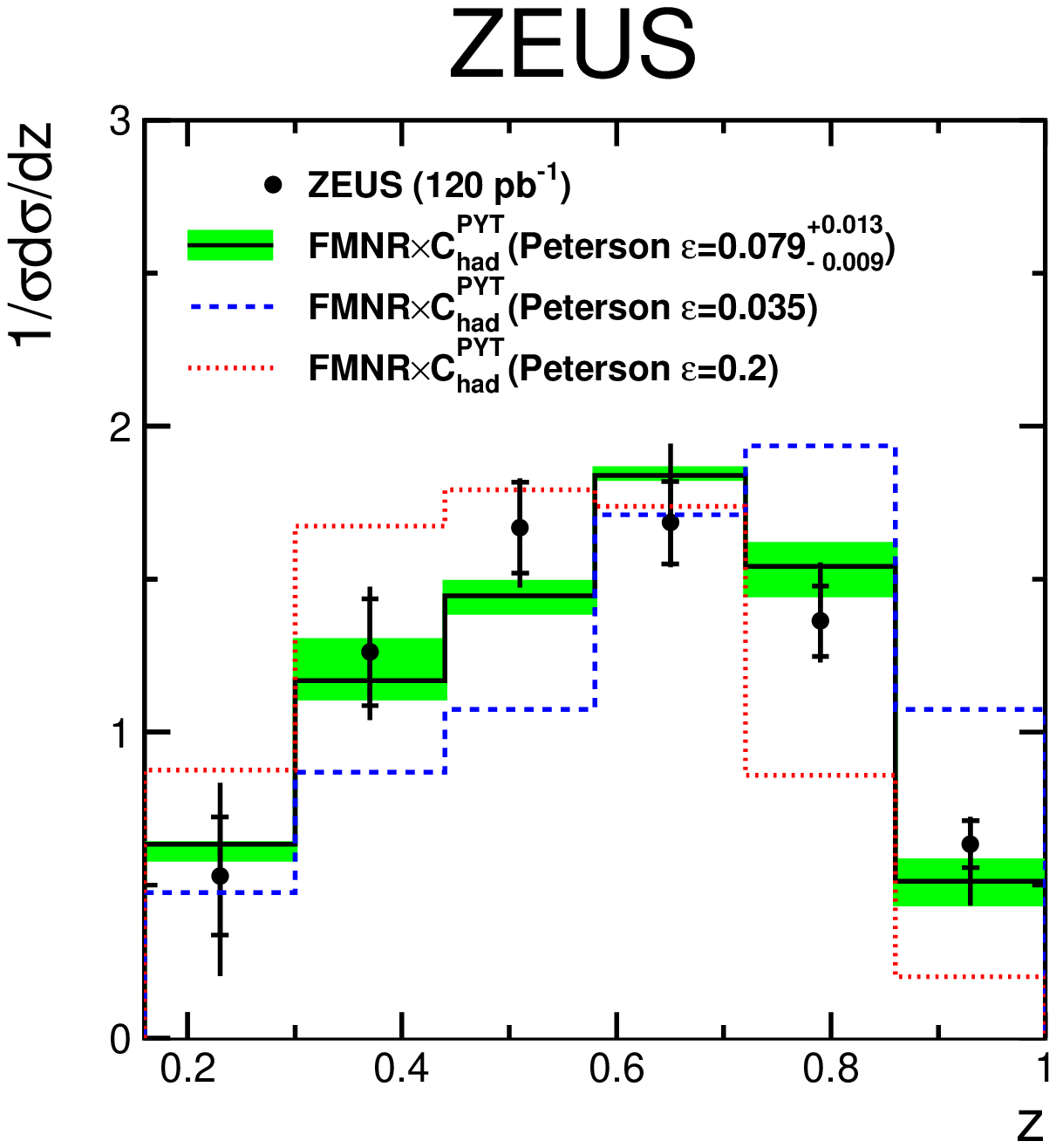}}
\end{center}
\vspace*{-1.0cm}
\caption[*]{Charm fragmentation function in transition to
$D^{*+}$ for the ZEUS data compared
to (left) measurements
in $e^+e^-$ annihilations
and (right) predictions of
FMNR$\times C^{\rm PYT}_{\rm had}$.}\label{Fig:zeus_cfragm}
\end{figure}

The measured fragmentation function
is compared in Fig.~\ref{Fig:zeus_cfragm}(left) with previous measurements
from Belle~\cite{belle_cfragm}, CLEO~\cite{cleo_cfragm}
and ALEPH~\cite{aleph_cfragm}.
For shape comparison, the data sets were normalised to
$1/$(bin width) for $z>0.3$.
The Belle and CLEO data are measured at a centre-of-mass energy of
$\sim10.5\,$GeV, whereas the ALEPH data were taken at $91.2\,$GeV.
The corresponding scale of the ZEUS data is given by twice the average
transverse energy of the jet, $23.6\,$GeV.
The ZEUS data in  Fig.~\ref{Fig:zeus_cfragm}(left) are shifted somewhat 
to lower values of $z$ compared to the CLEO and BELLE data with the ALEPH data
even lower that is consistent with expectations
from scaling violations in QCD.

The ZEUS data were compared with fragmentation models
implemented in the leading-logarithmic Monte Carlo (MC) program
PYTHIA~\cite{pythia}.
The LUND string fragmentation model~\cite{lund}
modified for heavy quarks~\cite{bowler} gives a reasonable
description of the data.
The PYTHIA predictions obtained using
the Peterson fragmentation function~\cite{peterson}
was fit to the data via a $\chi^2$-minimisation procedure
to determine the best value of the parameter $\epsilon$.
The result of the fit is
$\epsilon=0.062\pm0.007^{+0.008}_{-0.004}$.
The result is in reasonable agreement with the default value used in
PYTHIA (0.05),
with the value measured by the H1 collaboration in deep inelastic
scattering ($0.061^{+0.011}_{-0.009}$)~\cite{h1_cfragm},
and with the value 0.053 obtained in the leading-logarithmic
fit~\cite{oleari}
to the ARGUS data~\cite{argus}.

The data were also compared with the next-to-leading-order (NLO) QCD
predictions from Frixione et al. (FMNR)~\cite{fmnr}.
The predictions with the parton-level jets were translated
to the predictions with the hadron-level jets
using the hadronisation correction factors, $C^{\rm PYT}_{\rm had}$,
obtained with the PYTHIA MC.
The result of varying $\epsilon$ in the Peterson function
for the predictions of FMNR$\times C^{\rm PYT}_{\rm had}$
is shown in the Fig.~\ref{Fig:zeus_cfragm}(right).
The fit result, $\epsilon=0.079\pm0.008^{+0.010}_{-0.005}$,
exceeds the value $0.035$ obtained from the NLO fit~\cite{oleari}
to the ARGUS data~\cite{argus}.
The fit of the FMNR$\times C^{\rm PYT}_{\rm had}$ predictions
with the fragmentation
function from Kartvelishvili et al.~\cite{kartve}
yielded the value of the free parameter
$\alpha=2.67\pm0.18^{+0.17}_{-0.25}$.

\section{Production of excited charm and charm-strange mesons}

The first study of excited charm and charm-strange mesons produced
in $ep$ collisions
was restricted to decays,
for which significant signals were identified:
\begin{eqnarray*}
D_1(2420)^0\, &\rightarrow& D^{*+}\pi^-,\\
D_2^{*}(2460)^0\, &\rightarrow&  D^{*+}\pi^-,D^{+}\pi^-,\\
D_{s1}(2536)^+ &\rightarrow& D^{*+}K^0_S,D^{*0}K^+.
\end{eqnarray*}
The measurement was performed in the full kinematic range of $Q^2$.
The $D^{*+}$ mesons were identified using the decay channel
$D^{*+}\rightarrow D^0\pi^{+}$ following by either
$D^0\rightarrow K^-\pi^+$ or $D^0\rightarrow K^-\pi^+\pi^+\pi^-$
decay.
The $D^+$ mesons were reconstructed in the decay
$D^+\rightarrow K^-\pi^+\pi^+$.
The $D^{*0}$ mesons were tagged in the decay to a $D^0$
and undetected neutrals following by the $D^0\rightarrow K^-\pi^+$
decay.

To extract the $D^0_1$ and $D^{*0}_2$ yields and properties,
a minimal $\chi^2$ fit was performed using simultaneously
the $M(D^+\pi^-)$ distribution and
the $M(D^{*+}\pi^-)$ distributions in four helicity intervals.
The helicity angle ($\alpha$) is defined as the angle between
the momenta of the additional pion and the pion from the $D^{*+}$
decay in the $D^{*+}$ rest frame.
The helicity angular distribution can be parametrised as
\begin{equation}
\frac{dN}{d\cos\alpha}\propto1+h\cos^2\alpha,
\label{eq:cosgen}
\end{equation}
where $h$ is the helicity parameter.
Heavy Quark Effective Theory~\cite{hqet}
(HQET) predicts $h=3$ ($h=0$) for the $1^+$ state
from the $j=3/2$ ($j=1/2$) doublet, and $h=-1$ for the $2^+$
state from the $j=3/2$ doublet.
Only $D$-wave decays are allowed for the members of the $j=3/2$ doublet;
therefore they are supposed to be narrow. On the other hand, the members
of the $j=1/2$ doublet decay through $S$-wave only and therefore are expected
to be broader~\cite{hqet2}.
Due to the finite charm quark mass
a separation of the two doublets is only an approximation and
amplitudes of two observable states with $J^P=1^+$
can be mixtures of $D$- and $S$-wave amplitudes.

The measured masses of the $D^0_1$ and $D^{*0}_2$ are
in reasonable agreement with the world average values~\cite{pdg08}.
The measured $D^0_1$ width is
$\Gamma(D^0_1)=53.2\pm7.2^{+3.3}_{-4.9}\,$MeV
which is above
the world
average value $20.4\pm1.7\,$MeV~\cite{pdg08}.
The observed difference can be a consequence of
differing production environments.
The $D^0_1$ width can have a sizeable contribution
from the broad $S$-wave decay even if
the $S$-wave admixture is small~\cite{sdmix}.
A larger $S$-wave admixture
at ZEUS
with respect to that in measurements with restricted
phase space,
which can suppress production
of the broad state,
could explain why the measured $D^0_1$ width is
larger than the world average value.

\begin{figure}[!ht]
\begin{center}
\centerline{\epsfxsize=2.7in\epsfbox{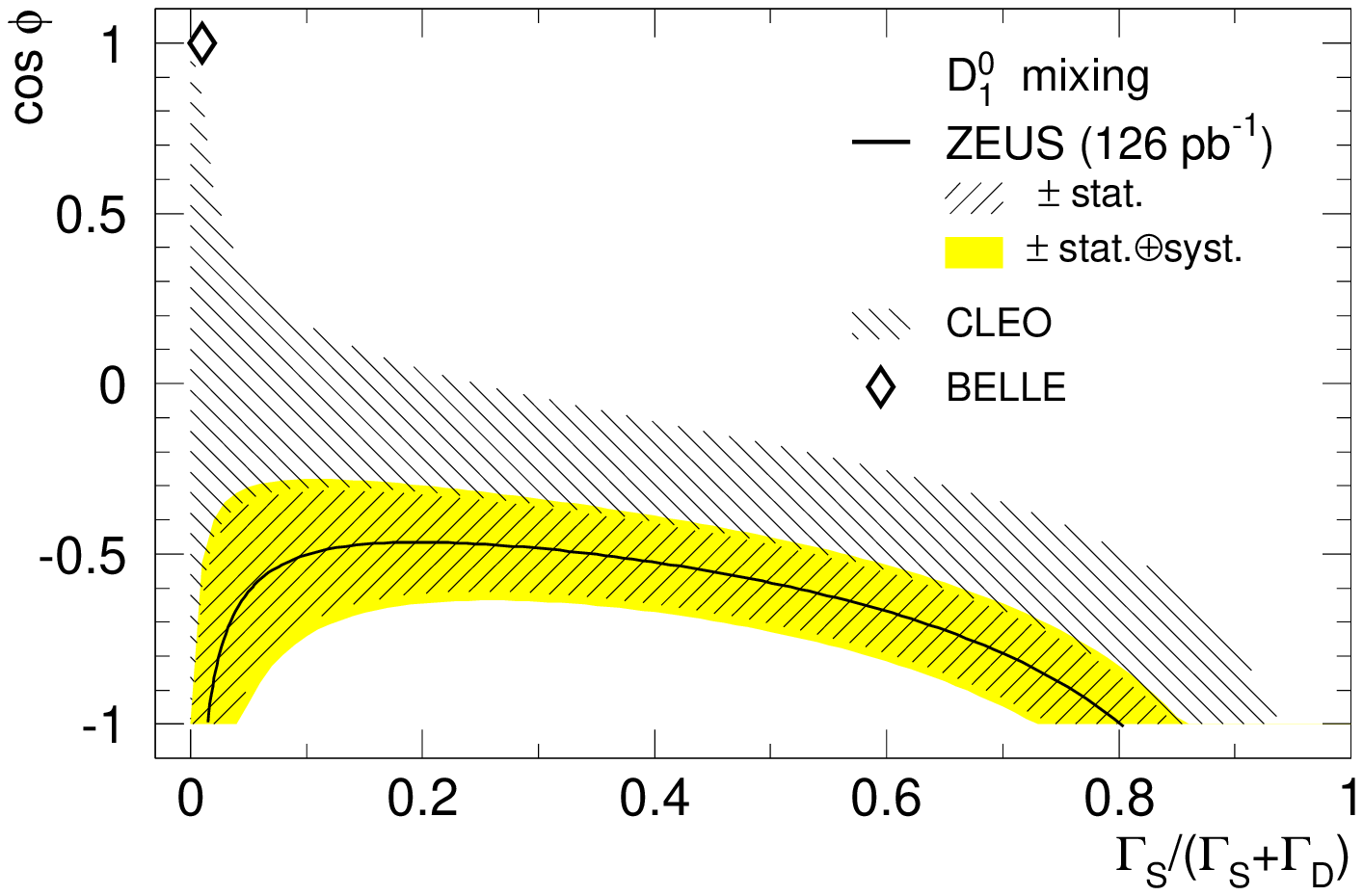}
\epsfxsize=2.7in\epsfbox{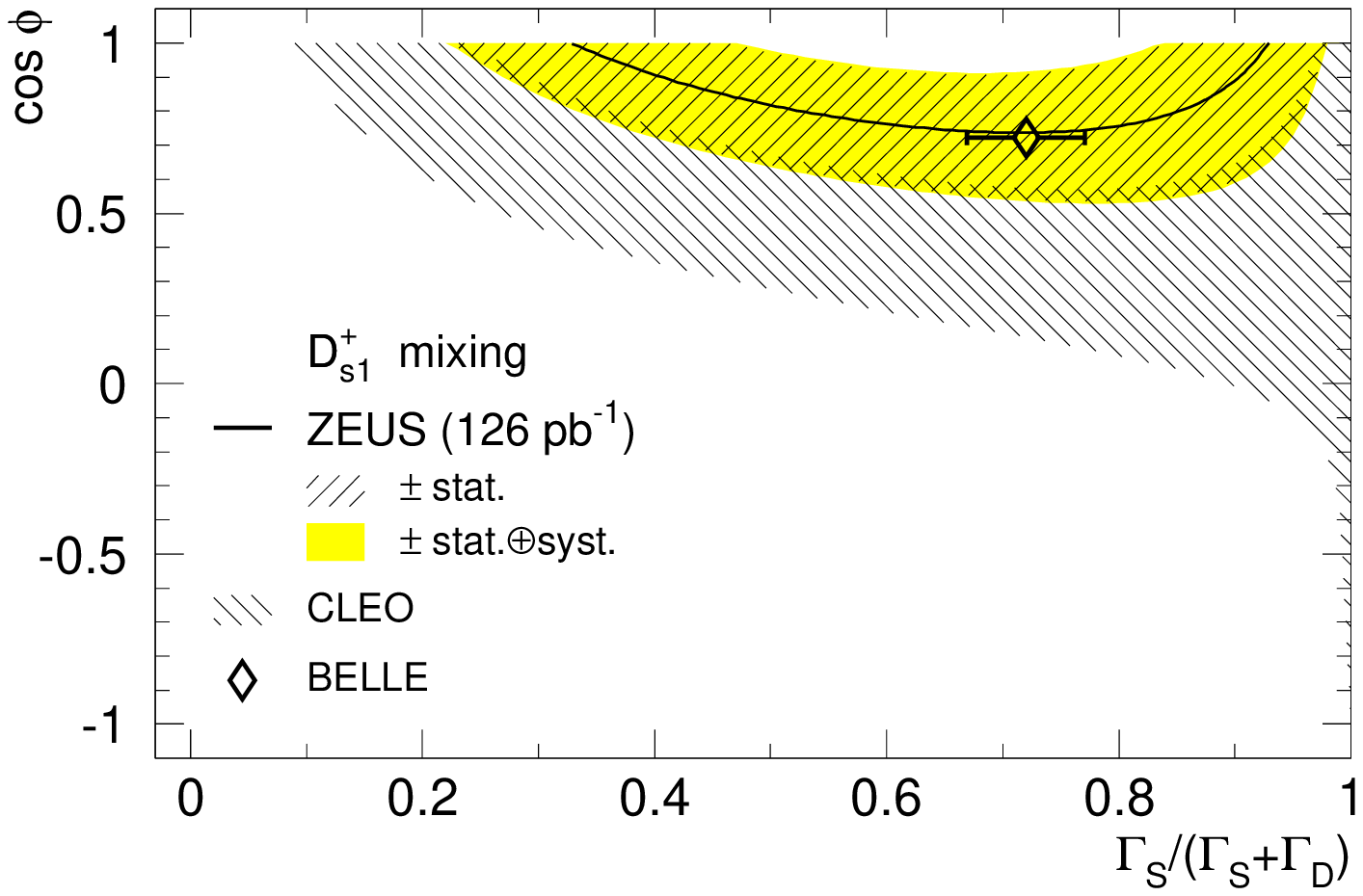}}
\end{center}
\vspace*{-1.0cm}
\caption[*]{Cosine of the relative phase of $S$- and $D$-wave
amplitudes versus $r=\Gamma_S/(\Gamma_S + \Gamma_D)$ in
(left) the $D_{1}(2420)^0\rightarrow D^{*+}\pi^-$ decay
and (right) the $D_{s1}(2536)^+\rightarrow D^{*+}K^0_S$ decay.}
\label{Fig:sdmix}
\end{figure}

The measured $D^0_1$ helicity parameter is
$h(D^0_1)=5.9^{+3.0+2.4}_{-1.7-1.0}$
which is inconsistent with the prediction
for a pure $S$-wave decay of the $1^+$ state, $h=0$. It is
consistent
with the prediction
for a pure $D$-wave decay, $h=3$.
In the general case of $D$- and $S$-wave mixing,
the helicity angular distribution form of the $1^+$ state is:
\begin{equation}
\frac{dN}{d\cos\alpha}\propto r + (1-r)(1+3\cos^2\alpha)/2
+\sqrt{2 r (1-r)} \cos\phi (1-3\cos^2\alpha),
\label{eq:cosmix}
\end{equation}
where $r=\Gamma_S/(\Gamma_S+\Gamma_D)$,
$\Gamma_{S/D}$ is the $S$-$/D$-wave partial width and $\phi$ is the relative
phase between the two amplitudes.
Using
Eqs.~(\ref{eq:cosgen}) and~(\ref{eq:cosmix}),
$\cos\phi$ can be expressed in terms of $r$
and the measured value of the helicity parameter, $h$: 
\begin{equation}
\cos \phi = \frac{(3-h)/(3+h)-r}{2\sqrt{2r(1-r)}}.
\label{eq:cosphi}
\end{equation}
Figure~\ref{Fig:sdmix}(left) compares
with previous measurements
the range restricted by the measured $h(D^0_1)$ value and
its uncertainties
in a plot of $\cos\phi$
versus $r$.
The ZEUS range has a marginal overlap with that restricted by the CLEO
measurement of  $h(D^0_1)=2.74^{+1.40}_{-0.93}$~\cite{cleo_sdmix}.
BELLE performed a three-angle analysis and measured
both the $\cos\phi$ and $r$ values~\cite{belle_sdmix}.
The BELLE measurement, which suggested a very small admixture of $S$-wave
to the $D_{1}(2420)^0\rightarrow D^{*+}\pi^-$ decay and
almost zero phase between two amplitudes, is outside
the ZEUS range;
the difference between the two measurements,
evaluated with Eq.~(\ref{eq:cosphi}),
is $\sim2$ standard deviations.

A signal from 
the $D^+_{s1} \rightarrow D^{*0}K^+$ decay
was observed in the $M(D^{0}K^+)$ distribution with an average
negative shift
of $142.4\pm0.2\,$MeV with respect to the nominal
$D^+_{s1}$ mass.
To extract the $D^+_{s1}$ yields and properties,
an unbinned likelihood fit was performed using simultaneously
values of $M(D^{0}K^+)$, $M(D^{*+}K^0_S)$, and $\cos(\alpha)$
for $D^{*+} K^0_S$ combinations,
with
the helicity angle defined as the angle between
the momenta of $K^0_S$ and the pion from the $D^{*+}$ decay
in the $D^{*+}$ rest frame.

The measured $D^+_{s1}$ mass is
in good agreement with the world average values~\cite{pdg08}.
The measured $D^+_{s1}$ helicity parameter is
$h(D^+_{s1})=-0.74^{+0.23+0.06}_{-0.17-0.05}$.
The measured $h$ value is inconsistent with the prediction
for a pure $D$-wave decay of the $1^+$ state, $h=3$, and is
barely consistent with
the prediction
for a pure $S$-wave decay, $h=0$.
Figure~\ref{Fig:sdmix}(right) shows
a range, restricted by the measured $h(D^+_{s1})$ value and
its uncertainties,
in a plot of $\cos\phi$
versus $r=\Gamma_S/(\Gamma_S + \Gamma_D)$ (Eq.~\ref{eq:cosphi}).
The measurement suggests a significant contribution
of both $D$- and $S$-wave amplitudes
to the $D_{s1}(2536)^+\rightarrow D^{*+}K^0_S$ decay.
The ZEUS range agrees with that restricted by the CLEO
measurement of  $h(D^+_{s1})=-0.23^{+0.40}_{-0.32}$~\cite{cleo_ds1}
and with the
BELLE three-angle measurement of
both $\cos\phi$ and $r$ values~\cite{belle_ds1}.

The measured yields
were converted to
the fragmentation fractions
$f(c\rightarrow D^0_1)=3.5\pm0.4^{+0.4}_{-0.6}\,\%$,
$f(c\rightarrow D^{*0}_2)=3.8\pm0.7^{+0.5}_{-0.6}\,\%$ and
$f(c\rightarrow D^+_{s1})=1.11\pm0.16^{+0.08}_{-0.10}\,\%$.
The fractions are consistent with those obtained
in $e^+e^-$ annihilations.
The measured ratios of the dominant $D^{*0}_2$ and $D^+_{s1}$
branching fractions are
$$\frac{{\cal B}_{D_2^{*0} \rightarrow D^+ \pi^-}}
{{\cal B}_{D_2^{*0} \rightarrow D^{*+} \pi^-}}=
2.8\pm0.8^{+0.5}_{-0.6}~~,~~
\frac{{\cal B}_{D_{s1}^+ \rightarrow D^{*0} K^+}}
{{\cal B}_{D_{s1}^+ \rightarrow D^{*+} K^0}}=
2.3\pm0.6\pm0.3$$
in agreement with the world average values~\cite{pdg08}.

No
radially excited $D^{*\prime+}$ meson,
reported by DELPHI~\cite{delphi_dsprime},
was observed.
An upper limit,
stronger than that obtained by OPAL~\cite{opal_dsprime},
was set on the product of the fraction of $c$ quarks
hadronising as a $D^{*\prime +}$ meson and the branching fraction
of the $D^{*\prime +}\rightarrow D^{*+}\pi^+\pi^-$ decay
in the range of the $D^{*\prime +}$ mass from $2.59$ to $2.69\,$GeV:
$$f(c\rightarrow D^{*\prime +}) \cdot {\cal B}_{D^{*\prime +} \rightarrow D^{\ast +} \pi^+ \pi^-} < 0.4 \%~~(95\%~~\rm{C.L.}).$$



\begin{footnotesize}




\begin{thebibliography}{99}
\bibitem{url} Slides: \\ 
\verb$http://indico.cern.ch/contributionDisplay.py?contribId=142&sessionId=5&confId=53294$
\bibitem{ff_php} ZEUS Coll., S.~Chekanov {\it et~al.}, Eur. Phys. J. {\bf C44} 351 (2005).
\bibitem{ff_dis} ZEUS Coll., S.~Chekanov {\it et~al.}, JHEP {\bf 07} 074 (2007).
\bibitem{zeus_func} ZEUS Coll., S.~Chekanov {\it et~al.}, JHEP {\bf 04} 082 (2009).
\bibitem{zeus_excd} ZEUS Coll., S.~Chekanov {\it et~al.}, Eur. Phys. J. {\bf C60} 25 (2009).
\bibitem{belle_cfragm} Belle Coll., R.~Seuster {\it et~al.}, Phys. Rev. {\bf D73} 032002 (2006).
\bibitem{cleo_cfragm} CLEO Coll., M.~Artuso {\it et~al.}, Phys. Rev. {\bf D70} 112001 (2004).
\bibitem{aleph_cfragm} ALEPH Coll., R.~Barate {\it et~al.}, Eur. Phys. J. {\bf C16} 597 (2000).
\bibitem{pythia} T.~Sj\"{o}strand, Comp. Phys. Comm. {\bf 135} 238 (2001).
\bibitem{lund}
B.~Anderson {\it et al.}, Phys. Rep. {\bf 97} 31 (1983).
\bibitem{bowler}
X.~Artru and G.~Mennessier, Nucl. Phys. {\bf B70} 93 (1974);\\
M.~G.~Bowler, Z. Phys. {\bf C11} 169 (1981).
 
\bibitem{peterson}
C.~Peterson {\it et al.}, Phys. Rev. {\bf D27} 105 (1983).

\bibitem{h1_cfragm} H1 Coll., F.D.~Aaron {\it et~al.}, Eur. Phys. J. {\bf C59} 589 (2009).

\bibitem{oleari}
P.~Nason and C.~Oleari, Nucl. Phys. {\bf B565} 245 (2000).

\bibitem{argus}
ARGUS Coll., H.~Albrecht {\it et al.}, Z. Phys. {\bf C52} 353 (1991).

\bibitem{fmnr} S.~Frixione et al., Phys. Lett. {\bf B348} 633 (1995);\\
S.~Frixione, P.~Nason and G.~Ridolfi, Nucl. Phys. {\bf B454} 3 (1995).

 \bibitem{kartve}
V.G~Kartvelishvili, A.K.~Likhoded and V.A.~Petrov, Phys. Lett. {\bf B78} 615 (1978).

\bibitem{hqet} N.~Isgur and M.B.~Wise, Phys. Lett. {\bf B232} 113 (1989);\\
M.~Neubert, Phys. Rev. {\bf A245} 259 (1994).

\bibitem{hqet2} N.~Isgur and M.B.~Wise, Phys. Rev. Lett. {\bf 66} 1130 (1991);\\
J.L.~Rosner, Comm. Nucl. Part. Phys. {\bf 16} 109 (1986).

\bibitem{pdg08} C.~Amsler {\it et~al.} (Particle Data Group), Phys. Lett. {\bf B667} 1 (2008).


\bibitem{sdmix} M.-Lu, M.B.~Wise and N.~Isgur, Phys. Rev. {\bf D45} 1553 (1992);\\
A.F.~Falk and M.E.~Peskin, Phys. Rev. {\bf D49} 3320 (1994).

\bibitem{cleo_sdmix} CLEO Coll., P.~Avery {\it et~al.}, Phys. Lett. {\bf B331} 236 (1994);\\
Erratum-ibid {\bf B342} 453 (1995).

\bibitem{belle_sdmix} Belle Coll., K.~Abe {\it et~al.}, Phys. Rev. {\bf D69} 112002 (2004).

\bibitem{cleo_ds1} CLEO Coll., J.P.~Alexander {\it et~al.}, Phys. Lett. {\bf B303} 377 (1993).

\bibitem{belle_ds1} Belle Coll., V.~Balagura {\it et~al.}, Phys. Rev. {\bf D77} 032001 (2008).

\bibitem{delphi_dsprime} DELPHI Coll., P.~Abreu {\it et~al.}, Phys. Lett. {\bf B426} 231 (1998).

\bibitem{opal_dsprime} OPAL Coll., G.~Abbiendi {\it et~al.}, Eur. Phys. J. {\bf C20} 445 (2001).

\end{thebibliography}
%

\end{footnotesize}


\end{document}